\documentclass[%
 reprint,
 amsmath,amssymb,
 aps,
]{revtex4-2}

\usepackage{graphicx}
\usepackage{dcolumn}
\usepackage{bm}

\begin{document}

\preprint{APS/123-QED}


\author{Dennis Scheidt}
\author{Pedro A. Quinto Su}%
 \email{Pedro.quinto@nucleares.unam.mx}
\affiliation{%
 Instituto de Ciencias Nucleares, Universidad Nacional Aut\'onoma de M\'exico, \\ Apartado Postal 70-543, 04510, Cd. Mx., M\'exico
}%


\date{\today}
\title{Errors in single pixel photography emerging from light collection limits by the bucket detector}

\begin{abstract}
In single pixel photography an image is sampled with a programmable optical element like a digital micromirror array or a spatial light modulator that can project an orthogonal base. The light reflected or diffracted is collected by a lens and measured with a photodiode (bucket detector).
In this work we demonstrate that single pixel photography that uses sampling bases with non-zero off-diagonal elements (i.e. Hadamard), can be susceptible to errors that emerge from the relative size of the bucket detector area compared with the spatial spread of the Fourier spectrum of the base element that has the the highest spatial frequency. 
Experiments with a spatial light modulator and simulations using a Hadamard basis show that if the bucket detector area is smaller than between $50-75\%$ of the maximum area spanned by the projected spectrum of the measurement basis, the reconstructed photograph will exhibit cross-talk with the effective phase of the optical system. The phase can be encoded or errors can be introduced in the optical system to demonstrate this effect
\end{abstract}


\maketitle
\section{Introduction}
Single pixel photography was introduced in 2006 by Baraniuk and coworkers \cite{Duarte}, where they demonstrated the ability to capture an image with a single photodetector, in contrast with regular digital cameras that use CCD or CMOS detectors that contain millions of individual sensors or pixels. Their approach projected an image into a digital micromirror array (DMD) that activates groups of micromirrors representing elements of an orthogonal basis. The reflected light was collected by a lens and detected by a single photodiode (bucket detector) placed at the focus of the lens. In this way, by sampling all the basis elements and recording the intensity of each basis element, it is possible to reconstruct the image. This has the advantage that it enables to obtain images at wavelengths that are outside regular CCDs or CMOSs sensors. 
Furthermore, it was also shown that compressive sensing could be used to reconstruct an image by only sampling a subset of the vectors contained in the measurement basis \cite{Duarte, candestao}.

Later, similar approaches were used to measure the phase of laser fields using the same single pixel photography approach of sampling the field with an orthogonal basis combined with interferometry \cite{SinglePixel}, but with the difference that the intensities are measured at a single point (single pixel in a camera) or selecting the light from a small area with a spatial filter followed by a photodiode \cite{Zupancic:16}, in contrast with the bucket detector approach of single pixel photography.

In this work we investigate the effect that the size of the bucket detector has relative to the area spanned by the Fourier spectrum of the elements of the sampling base. We use a complete Hadamard basis to sample and reconstruct an image and show that errors like aberrations, and even phase components contained in the field can appear in the reconstructed image due to cross-talk that emerges from incomplete integration by the bucket detector of the light projected by the elements of the sampling base.

\section{Single Pixel Photography}
Lets consider a complex field $u$ that is represented by a vector with $N$ components. A photograph of the complex field will only include the intensity and the resulting image can be represented by a vector $x\propto |u|^2$. The image $|u|^2$ can be sampled by an orthogonal base $\Phi$ ($N\times N$ matrix), where the orthogonal ($1\times N$) vectors are the rows of that matrix. The projections of the image in each orthogonal base element are $y=\Phi u$, where $y$ is the vector of the integrated measured intensity for each vector that makes the orthogonal base.

Most single pixel photography implementations use the Hadamard basis $\Phi =H$ because it spans the full effective area of the programmable optical element (because it has no zeros), in contrast to the canonical basis represented by the identity matrix $I$ ($\Phi=I$). The Hadamard matrix has the following property $HH^T =H^2= NI$.
Many experiments divide the Hadamard matrix into a positive part and a negative part $H= H^+ -H^-$, so that $H^+ =(H+1)/2$ and $H^-=-(H-1)/2$ (both binary matrices containing zeros and ones). In this way $y=y^+ -y^-$ with $y^+=H^+ u$ and $y^-=H^- u$.  
The image is reconstructed with $x = H (y^+ - y^-) = H (H^+ u - H^- u) = H (H^+ - H^-) u = H H u=Nu$.

\subsection{Field and basis encoding}
Lets define the discrete complex light field at the surface of an SLM (or DMD) with spatial coordinates $(X,Y)$: $u(X,Y)=A(X,Y)e^{i\phi (X,Y)}$. The field is resized into a $\sqrt{N}\times \sqrt{N}$ square grid made of $N$ superpixels, where each superpixel is made of several pixels. The same procedure is done for the 1-d vector elements of the bases $\Phi_l^{\pm}$. In this way, the projection of the field on the a basis element $\Phi _l ^{\pm}$ is encoded with $\Phi _l ^{\pm}(X,Y)\odot u(X,Y)$, where $\odot$ is the element-wise multiplication. Since that new field is also encoded in the same area of the SLM it is also a function of the same spatial coordinates. 
 
The light diffracted by the new encoded field can be collected by a lens (focal length $f$) that focuses it on a photodiode (bucket detector) that integrates the power that base element $y_l ^{\pm}$. The focused intensity  pattern is the 2-dimensional Fourier transform of the field projected at the surface of the SLM:
\begin{multline}
    \mathcal{F}\{\Phi_l^{\pm} (X,Y)\odot u(X,Y)\}_{k_X,k_Y} =\\
    \int_{\Delta X} dX \int_{\Delta Y} dY \ \Phi_{l}^{\pm}(X,Y) u(X,Y) 
    e^{-2\pi i (k_X X+ k_Y Y) }  \ .
\end{multline}
where $\mathcal{F}$ denotes the 2 dimensional Fourier transform that is applied by the focusing lens with their respective frequency components $k_x = X/\lambda f$ and $k_y = Y / \lambda f$. $\Delta X$ and $\Delta Y$ represent the size of the square aperture at the SLM in $X$ and $Y$, respectively.\\
The intensity pattern at the focus is:
\begin{multline}
    y_l^{\pm} = \int_{\Delta X} \int_{\Delta Y} dX \ dY \ \Phi_{m}^{\pm}(X,Y) u(X,Y) \cdot \\
    \int_{\Delta X} \int_{\Delta Y} dX' \ dY' \ \Phi_{l}^{\pm *}(X',Y') u^{*}(X',Y') \cdot \\
    \iint dk_X \ dk_Y \ e^{-2\pi i\left(k_X(X-X') + k_Y(Y-Y') \right)} \ .
\end{multline}
Using $u(X,Y) = A(X,Y) e^{i\phi (X,Y)}$ yields:
\begin{multline}
    y_l^{\pm} = \int_{\Delta X} \int_{\Delta X} \int_{\Delta Y} \int_{\Delta Y} dX dX' dY dY'\ \Phi_{l}^{\pm}(X,Y)
     \cdot \\
     \Phi_{l}^{\pm}(X',Y')
    A(X,Y) A(X',Y') e^{i(\phi(X,Y) - \phi(X',Y'))} \cdot \\ 
    \iint dk_X dk_Y \  e^{-2\pi i\left(k_X(X-X') + k_Y(Y-Y') \right)} \ .
\end{multline}

\noindent The photodiode integrates the signal over the Fourier spectrum, yielding the measured power as: 
\begin{equation}
    y_l^{\pm} = \int dk_x \int dk_y |\mathcal{F}\{\Phi_l^{\pm} (X,Y)\odot u (X,Y)\}_{k_X,k_Y}|^2 \ . 
\end{equation}
\noindent 
Ideally the photodiode will collect the full spectrum, so that there is no power loss. 
\noindent Evaluating the frequency integrals ($k$) over the full spectrum yields:
\begin{equation}
    \iint_\infty dk_X dk_Y \  e^{-2\pi i\left(k_X(X-X') + k_Y(Y-Y') \right)} = \delta(X-X', Y - Y') \ 
\end{equation}
where $\delta $ is the Dirac delta function.
\noindent In this way, $y_l^{\pm}$ reduces to:
\begin{equation}
    y_l^{\pm} = \int_{\Delta X}\int_{\Delta Y} dX dY |\Phi_{l}^{\pm}(X,Y) A(X,Y)|^2 \ .
\end{equation}
which is what is expected for single pixel photographs: The measurements for each element of the basis integrate the intensities and there is no phase information. \\

\noindent Note, that $\int_{\Delta X} dX\int_{\Delta Y} dY|\Phi_m^{\pm}(X,Y) A(X,Y)|^2 = \int_{\Delta X} dX\int_{\Delta Y} dY\Phi_{l}^{\pm}(X,Y) |A(X,Y)|^2 = \Phi_l^{\pm}(X,Y) \cdot |A(X,Y)|^2$,  because the basis $\Phi^\pm=H^\pm$ has only entries of $0$ or $1$.
Thus, allowing to recover the image of photograph $|A|^2$ scaled by a constant using $x = H y =H(y^+- y^-)= H H |A|^2 = N |A|^2$.

In the case that the bucket detector does not integrate the full spectrum, then the integrals over $k$ in equation (4) are evaluated over arbitrary boundaries $A$ and $B$. The $k_X$ integral is $\int_{A}^B dk_X e^{-2\pi i k_X (X-X')} = -i \frac{e^{-2\pi i B (X-X')} - e^{-2\pi i A (X-X')} }{2 \pi (X - X')}$.
For the special case of a symmetric integration of the spectrum, that is $A = - B=r_k$, the expression simplifies to:
\begin{equation}
    \int_{-r_k}^{r_k} dk_X \ e^{-2\pi i\left(k_X(X-X') \right)} = 2r_k sinc \left( 2\pi r_k (X-X') \right)  \ .
\end{equation}
Note, that increasing $r_k$, the expression approaches the Dirac delta function. \\
Evaluating equation (4) in the case of a truncated integration over $(-r_k,r_k)$ for both $k_X , k_Y$T (square detector with area $4 r_k ^2$) reads: 
\begin{multline}
    y_l^{\pm} = \int_{\Delta X} \int_{\Delta X} dX dX' \int_{\Delta Y} \int_{\Delta Y} dY dY' \  \Phi_{l}^{\pm^2} A^2(X,Y) \cdot \\ 
   4r_k^2 e^{i(\phi(X,Y) - \phi(X',Y')} \cdot \\
   \text{sinc} \left( 2\pi r_k (X-X') \right) 
    \text{sinc} \left( 2\pi r_k (Y-Y') \right)\ .
\end{multline}
This expression is independent of the sampling basis (e.g. Gaussian, Hadamard).
As a consequence of eq. (8),  phase crosstalk $\exp \left(i (\phi(X,Y) - \phi(X',Y')) \right)$ emerges, resulting in phase induced artifacts in the reconstructed photograph. The exception would be a basis that only contains diagonal elements (e.g. canonical).\\

Next, we quantify the spatial size of the detector ($2r_k$) that is needed to avoid phase cross-talk in the reconstructed photograph. This is done as a function of the focal length of the lens, the size of encoded field and the maximum spatial frequency reached by the measuring base.

\subsection{Projected Pattern Size}

The maximum spatial range covered by the focused basis elements can be calculated for each direction using the 1d Fourier transform of the Hadamard elements, as the Fourier transform is separable for each dimension.
The pattern with the highest spatial frequency is that of a checkerboard with the squares of length $a$ so the maximum spatial frequency $f_{Xmax} =1/(2a)$.

So in one direction ($X$ or $Y$) this is represented by a pulse train containing a periodic square wave with a width of $a$ which has the length of an individual superpixel. 
The length of the pulse train is $L=\sqrt{N}a$.
The pulse train is described by \cite{goodman2005introduction} $f(X) = \left(\text{rect}(X/a) \otimes \text{comb}(X-2a) \right)\cdot\text{rect}(X/L)$, where rect(X) and comb(X) are the rectangular and Dirac comb function, respectively.
The resulting intensity is proportional to the square of the Fourier transform calculated in reference \cite{goodman2005introduction}: 
\begin{equation}
    I(f_X) \propto \sum_{m = -\infty}^{m = \infty} \left( \text{sinc}(m/2)\text{sinc}(La(f_X - f_{Xmax}) \right)^2 \ ,
\end{equation}
\noindent where $f_X = X/\lambda f$ and $m$ is an integer. 
The intensity is determined by the first sinc function sinc$(m/2)$ and decays rapidly with increasing $m$. We therefore define the extent of the spectrum with the position of the first observed peak, that occurs when the argument of the second sinc function is zero, at $f_X=f_{Xmax}$ so that $1/(2a)=X/(\lambda f)$. Hence the spatial dimension of $r_k$:
\begin{equation}
r_k = \lambda f /(2 a )\ .     
\end{equation}

\noindent Hence, in order to eliminate phase crosstalk, we need a detector with a spatial size of $2r_k$.

\begin{figure}
    \centering
    \includegraphics[width=0.45\textwidth]{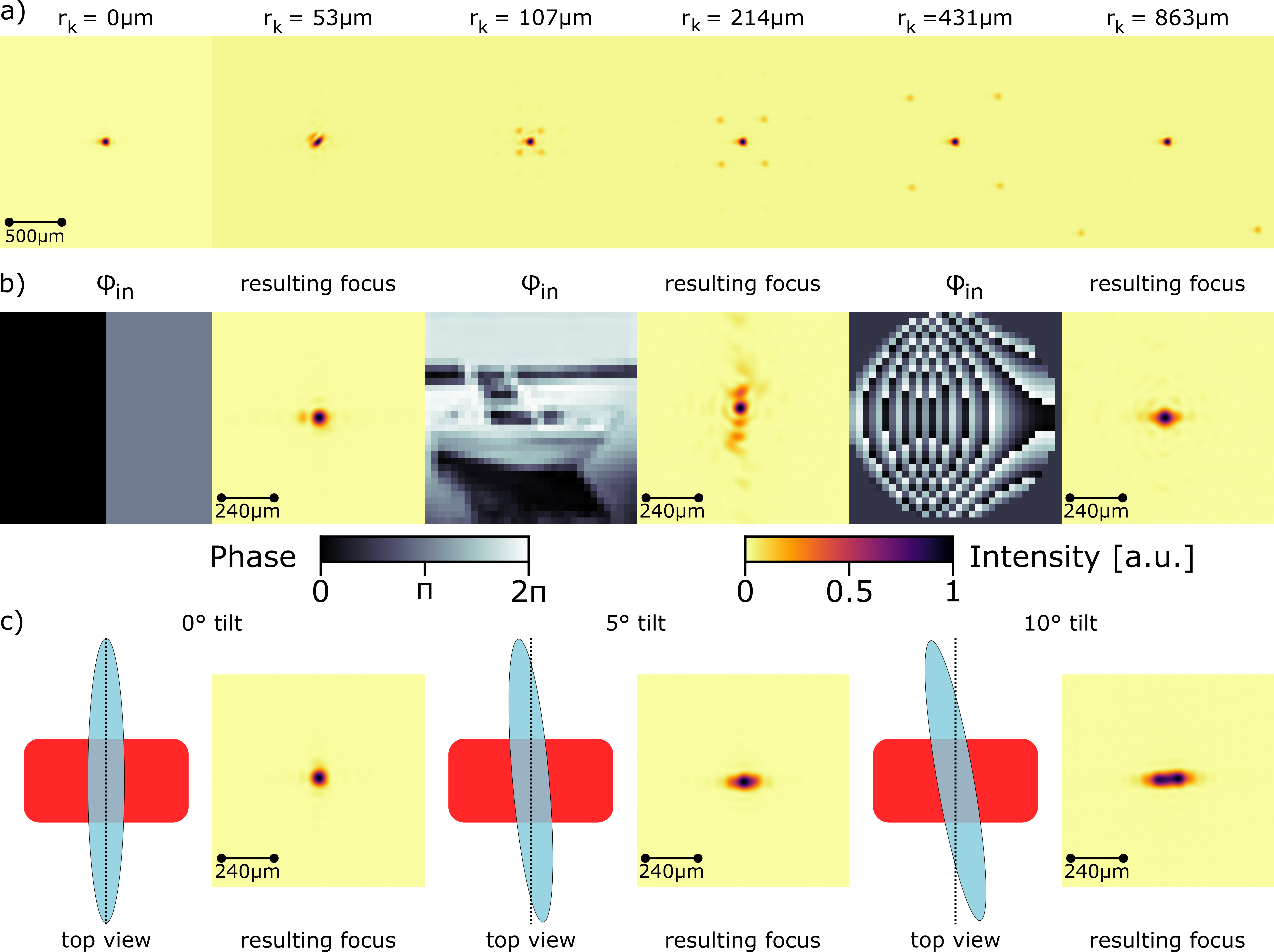}
    \caption{Spectral properties: (a) Extend of the Hadamard spectrum of different basis vectors according to equation (10). Frame width of 2 mm. (b) Aberrated focal spot due to different phases: vertical discontinuity, boat phase and tilt plus astigmatism ($\phi(\rho)=\text{mod}(4\pi Z_1^1(\rho)+ \pi Z_2^2(\rho), 2\pi)$). Frame width is 1 mm. (c) Aberrated focal spot due to tilting of the focusing lens. Frame width of 1 mm.}
    \label{fig1}
\end{figure}

\section{Experimental Setup}
The experimental setup is similar to the one described in \cite{Scheidt:23} and to the original of Baraniuk and coworkers \cite{Duarte} that used a DMD. In our case, the field is sampled with a SLM (Hamamatsu LCOS-SLM X10468) that has a pixel size of $20\,\mu$m. The light source is an expanded HeNe laser (wavelength $\lambda=633\,$nm).

The amplitude of the fields at the SLM has the shape of the expanded Gaussian or it is a $32 \times 32$ pixel image from the CIFAR-10 database \cite{cifar10}. An arbitrary phase defined in the same grid can be added to create a complex field. Both, amplitude and phase are encoded with a phase-only hologram \cite{Bolduc:13} and pixel cross-talk in the hologram is reduced by applying a Gaussian filter to decrease the effect of phase discontinuities \cite{cross_talk_holography}.

The field is encoded at the SLM in an area of $384\times 384$ pixels that makes a square aperture with a size of $7.68 \times 7.68\,$mm$^2$. 
So the $32\times 32\,$ field is resized to the square aperture resulting in superpixels that have a size of $240 \times 240\,\mu$m$^2$  ($12\times 12$ px$^2$). A Hadamard basis with 1024 elements ($1024\times 1024$ matrix) is used to sample the field. The positive and negative entries of the Hadamard basis are sampled separately ($H = H^+ + H^-$). Each element is also resized and reshaped in the same way as the field. In this way, the encoded projection of the field on a basis element is the encoded field basis element $\Phi _m ^{\pm} \odot u$ like in \cite{OurPaper}. 

Then, a telescope composed of two lenses with focal lengths of 50 cm and 15 cm, reduces the spatial size of the encoded field by a factor of $15/50$, so the square aperture size is now $2.304 \times 2.304\,$mm (effective SLM pixel size reduced to $6\,\mu$m). This is followed by a lens with $f=20\,$cm that focuses the diffracted light (first order, spatially filtered at the first lens) into a CCD camera.
We use a CCD camera because it enables us to integrate the signal across different regions of interest to simulate bucket detectors with different sizes in the range of [$3.6\times 3.6, 1440 \times 1440\mu$m$^2$]. The width of the Hadamard pattern with the maximum spatial frequency (checkerboard pattern) is $a=72\,\mu$m, so that $r_k=879\,\mu$m.\\

Figure 1(a) shows the different focused positive Hadamard basis elements $H^+_m$ for checkerboard-like patterns with increasing spatial frequencies. The last frame shows the projected intensity for the pattern with the highest spatial frequency $H^+_{34}$ that has the maximum spatial spread. We observe that the intensity pattern is made by five spots, one at the center and four near the corners. The distance from the center (horizontal/vertical) to one of the spots at the corners is $\approx 860\mu$m, which is consistent with the calculated $r_k$ of $879\,\mu$m.

Figure 1(b) shows the shape of the focused spots for different fields: One with a vertical phase discontinuity of $\pi$, another with a phase corresponding to a tilt and astigmatism  aberration with $\phi(\rho)=\text{mod}(4\pi Z_1^1(\rho)+ \pi Z_2^2(\rho), 2\pi)$ with normalized Zernike polynomials. The last case has an arbitrary phase with the shape of a boat. Finally, Fig. 1(c) depicts the focused spots of beams without an added phase where the focusing lens is twisted at different angles, generating an aberration that is a mixture of tilt and astigmatism. It is important to mention that as the center of the focused beam shifts slightly depending on the applied phase or twist of the lens, before starting the measurement the evaluated region of interest (ROI) in the camera is centered around the maximum of the disturbed focal spot.\\  

\subsection{Simulation}
The experiment is simulated using the experimental parameters and the discrete versions of equations (1-9). Where the Fourier Transform is replaced by a Discrete Fourier Transform (DFT). The main difference is that the integrals become sums and in the case of the integrals over the full frequency spectrum (eq.(5)) yield the Kronecker delta $\delta _{nm}$ instead of the Dirac delta function, yielding $y_l^{\pm}=\sum _{n,m}|\Phi _{nm} A_{nm}|^2$, where $l$ denotes the basis element and coordinates $(X,Y)$ are changed to indices $n,m$. Also in the case of not integrating over the whole spectrum, eq.(4) has to be discretized, which can be found in the supplementary document.
The code for the simulation is provided in \cite{supplement_code}.

\section{Results}

Figure 2(a) shows the effect of the detector size in simulations for an amplitude with an encoded car photograph and different phase distributions in each row: Constant phase, $\pi$ vertical discontinuity, boat image encoded in the base and an encoded aberration of tilt and astigmatism aberration like that of Fig. 1(b). 
Along each column the amount of the sampling of the Fourier space increases. In the case of a constant phase (first row), no difference in the detector size is noticeable. However, already the split phase case (second row) introduces a vertical line in the DC term that is also present in when sampling $25\%$ of the spectrum. The boat phase (third row) disappears at $50\%$ coverage. Finally (fourth row), the Zernike aberration is persistent even at $75\%$ and only disappears completely when the detector covers the full Fourier spectrum.

\begin{figure}
    \centering
    \includegraphics[width=0.4\textwidth]{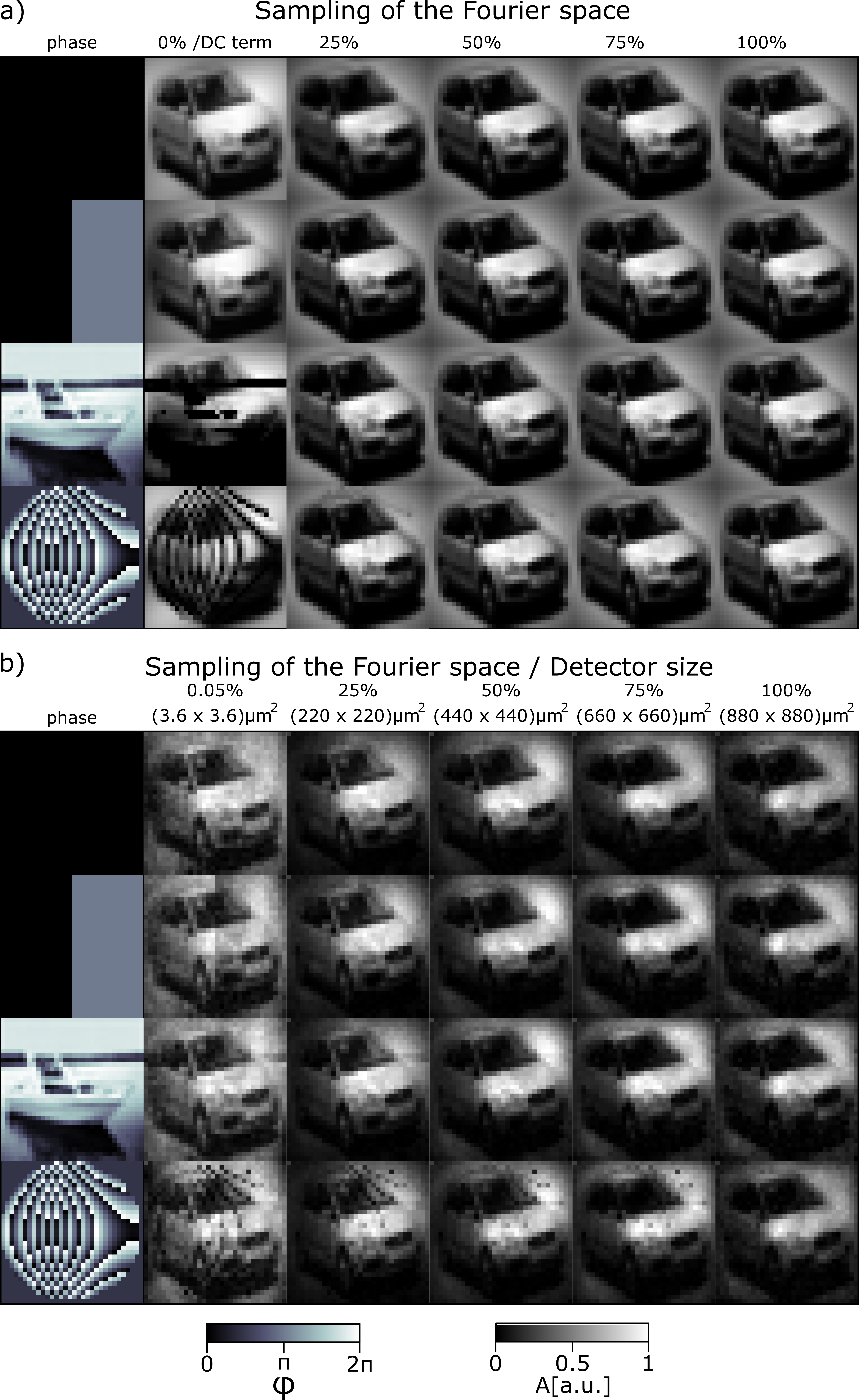}
    \caption{Sampling size of the Fourier spectrum with Single Pixel Imaging. (a)) Simulation and (b) experimental results dependent on the sampling size of the Fourier spectrum. Major effects of aberrations disappear when sampling over $50\%$ of the Fourier spectrum and completely disappear when the full spectrum is sampled.}
    \label{fig2}
\end{figure}

Figure 2(b) presents the results for the experiment using the same car amplitude with the same phase profiles encoded by the SLM. The columns feature different detector sizes that correspond to a certain amount of the Fourier spectrum. In all cases the results are similar to the simulation but there is more noise. For example, the most extreme case of the tilt and astigmatism aberration (fourth row) the phase has more contrast than in the simulation and it is still noticeable at $75\%$ of the integrated Fourier spectrum.

\vspace{0.2cm}
Figure 3 shows experimental single pixel photographs with no amplitude or phase encoding at the SLM (Fig. 3(a), only the reflected Gaussian beam) and with the car image encoded in the amplitude (Fig. 3(b)).
In both cases the rows represent experiment with the collection lens with different degrees of tilt: $0^{\circ}$, $5^{\circ}$ and $10^{\circ}$, while the detector size increases along the columns. 
For both cases, no change in the acquired image is apparent when the lens is perpendicular towards the optical axis -- that is, no aberrations are introduced into the field. However, already a slight tilt of $5^{\circ}$ is noticeable for small detector sizes $3.6 \mu$m and $90\mu$m as dark vertical lines appear close to the border. 
Increasing the tilt to $10^{\circ}$ introduces two slightly bent vertical lines on the right border for detector sizes that sample up to $25\%$ of the Fourier spectrum. At $50\%$ the recovered amplitude still features a shadow on the right side and is indistinguishable after sampling $75\%$ of the Fourier spectrum.

\begin{figure}
    \centering
    \includegraphics[width=0.4\textwidth]{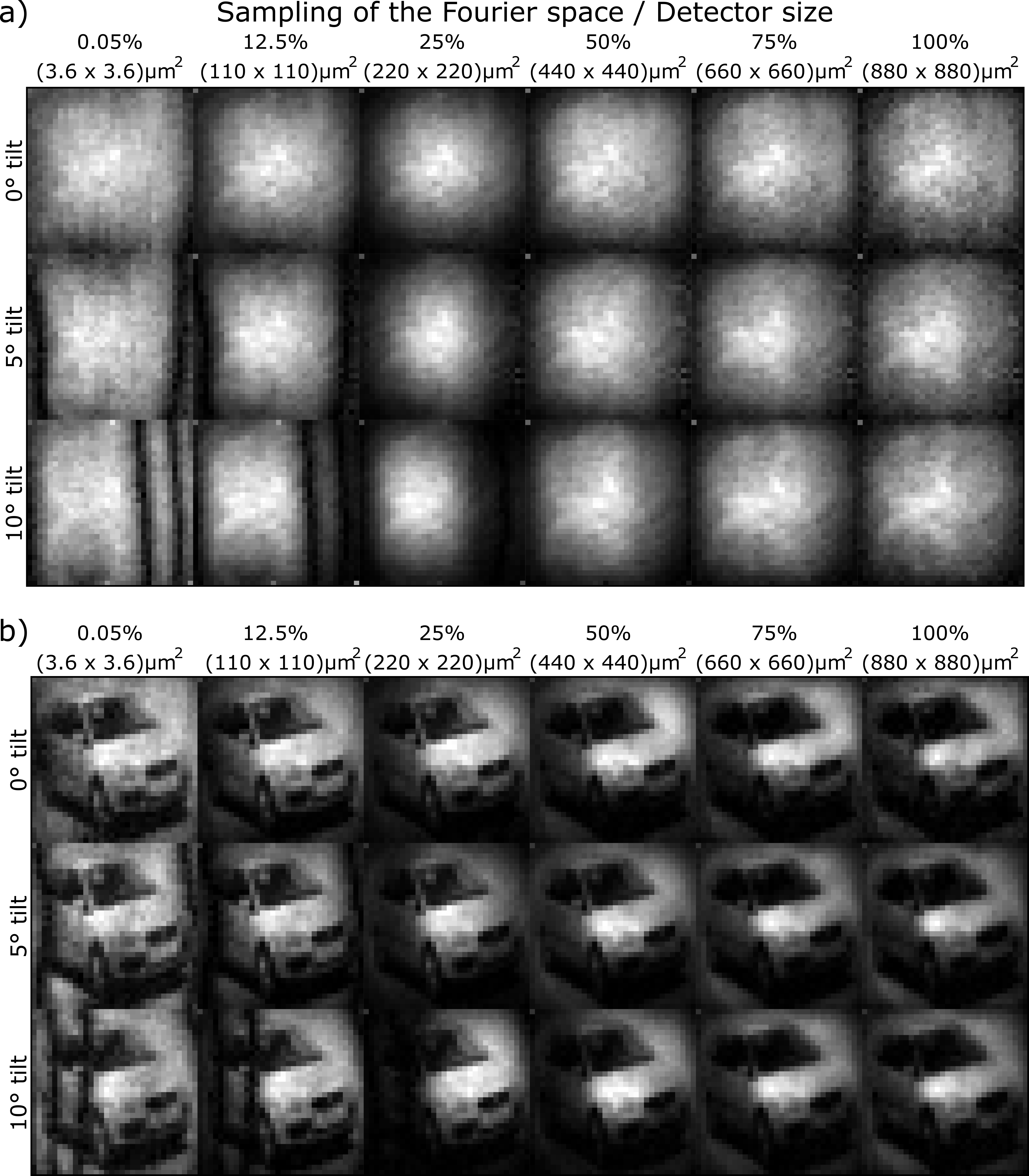}
    \caption{Single Pixel Imaging with a titled lens dependent on the detector size. a) No amplitude modulation, b) Amplitude modulated with a 'car' image. Data for $0^\circ$ tilt is the same as for the first row of Fig. 2b.}
    \label{fig3}
\end{figure}

\section{Conclusion}

We have shown the importance of the bucket detector size in single pixel photography as a function of the light projection system. In order to cover the full Fourier spectrum it is important that the detector has dimensions of $2r_k=\lambda f/2a$.
For a detector that has a size of less than between $50-75\%$ of $2r_k$, the phase distribution is visible in the reconstructed image. Conventional photodiodes can have large sizes on the order of 10\,mm, while fast photodiodes have smaller dimensions of a few hundreds of micrometers. An extreme case would be that of coupling the light to a single mode optical fiber. 
The effect of crosstalk with the phase could be very important in biological applications, such as image transmission through biological or random media \cite{Tajahuerce:14}. In this case light interacts with micro-structures, so that a randomized phase distribution can appear in the transmitted image.

\bibliography{bibliography}

\end{document}